\documentclass[twocolumn,aps,showpacs,prb,tightenlines,amsmath,amssymb]{revtex4}
\usepackage{graphicx}
\usepackage{amssymb}
\usepackage{slashed}
\usepackage{dcolumn}
\usepackage{amsmath}
\usepackage{bm}% bold math
\usepackage{colordvi}
\usepackage{mathrsfs}
\makeatletter

\newcommand{\Rmnum}[1]{\expandafter\@slowromancap\romannumeral #1@}
\makeatother

\begin{document}

\title{Comment on ``Nonlinear electromagnetic response and Higgs-mode excitation in BCS superconductors with impurities''}

\author{F. Yang}
\email{yfgq@mail.ustc.edu.cn}

\affiliation{Hefei National Laboratory for Physical Sciences at
Microscale, Department of Physics, and CAS Key Laboratory of Strongly-Coupled
Quantum Matter Physics, University of Science and Technology of China, Hefei,
Anhui, 230026, China}

\author{M. W. Wu}
\email{mwwu@ustc.edu.cn.}

\affiliation{Hefei National Laboratory for Physical Sciences at
Microscale, Department of Physics, and CAS Key Laboratory of Strongly-Coupled
Quantum Matter Physics, University of Science and Technology of China, Hefei,
Anhui, 230026, China}

\date{\today}

\pacs{74.40.Gh, 74.25.Gz, 74.25.N-}
%74.40.Gh Nonequilibrium superconductivity
%74.25.Gz Optical properties
%74.25.N- Response to electromagnetic fields

\maketitle

Very recently, by separately using Eilenberger equation
and diagrammatic formalism, Silaev\cite{CO} studied the Higgs-mode excitation
(i.e., amplitude fluctuation $\delta|\Delta|$ of the order parameter $\Delta$)
in the
presence of the scattering. This is the paper that for the first time the scattering influence on the optical properties is calculated
within the Eilenberger equation.\cite{Eilen} Unfortunately, in that paper, the main conclusions that the Higgs-mode generation is zero without impurity and the Higgs mode is not sensitive to disorder, are not correct. The former arises from an error in the calculation, while the latter is due to the improper approximation of the used approach. In the following, we address the reasons in detail.   

Firstly, in the paper by Silaev, the
claimed conclusion that the Higgs-mode generation is zero without impurity is incorrect. This can be easily seen from the Ginzburg-Landau theory\cite{G1,Ginzburg} which is established in the clean limit. Specifically, in consideration of the vector potential alone, the Ginzburg-Landau Lagrangian is given by\cite{G1}
\begin{equation}
  L=c|i\partial_t\Psi|^2\!-\!\Big[a|\Psi|^2\!+\!\frac{b}{2}|\Psi|^4\!+\!\frac{1}{4m}|(-i{\bm \nabla\!-\!2e{\bf A}})\Psi|^2\Big],
\end{equation}
with $a$, $b$ and $c$ being the expansion parameters in the Ginzburg-Landau theory. Then, in the long-wave limit, under the expansion of the Landau order parameter $\Psi=\Psi_0+\delta|\Psi(t)|$, one finds the equation of motion 
\begin{equation}
(c\partial_t^2-2a)\delta|\Psi|=-\frac{(2eA)^2\Psi_0}{4m},  
\end{equation}
exactly telling the finite second-order electromagnetic response of the Higgs mode in the clean limit. Hence, the claimed conclusion above by Silaev is incorrect.

The study using Eilenberger equation should have recovered this finite response, since the Eilenberger equation can reduce to Ginzburg-Landau equation near the critical temperature (refer to Appendix~\ref{A_GC}). However, Silaev in Ref.~\onlinecite{CO} overlooked the lower limit of the summation for the Matsubara frequency (i.e., $\omega_n>0$). Such an error leads to the vanishing electromagnetic effect in the derivation of the Ginzburg-Landau equation from Eilenberger equation (refer to Appendix~\ref{A_GC}), and this is exactly the reason that Silaev in his paper reported the vanishing second-order electromagnetic response of the Higgs mode in the clean limit. In fact, considering the lower limit of the summation for the Matsubara frequency (i.e., $\omega_n>0$), Eq.~(60) in Ref.~\onlinecite{CO}, i.e., the second-order electromagnetic response of the Higgs mode, is finite. 

Furthermore, it is also noted that the electromagnetic effect in the Eilenberger equation is incomplete. Specifically, the Eilenberger equation with vector potential ${\bf A}$ alone is not gauge invariant.\cite{gi0,gi1} 
It is well known that the gauge invariance is the basic character of the electromagnetic
field. The absence of the gauge invariance indicates incomplete
electromagnetic effect. This incompleteness in Eilenberger equation can be clearly seen as follows. 

In the electromagnetic response, it is established that
the superconductors directly respond to vector potential ${\bf A}$ (Meissner and Ginzburg-Landau effects) in addition to electric field ${\bf E}=-\partial_{\bf R}\phi-\partial_T{\bf A}$,  differing from the normal metals that respond to ${\bf E}$ solely. Therefore, to elucidate complete picture for optical response of superconductors, both effects by ${\bf E}$ and directly by ${\bf A}$ must be included.
In the non-gauge-invariant Eilenberger\cite{Eilen,CO} equation, although the paramagnetic effect of ${\bf A}$ (i.e., $H_p={\bf p}\cdot{e}{\bf A}/m$) on anomalous Green function is involved, making this equation well tailored to handle stationary magnetic response\cite{Ba7,Ba8} and derive the Ginzburg-Landau equation,\cite{Ba20} the diamagnetic effect of ${\bf A}$ [i.e., $H_d=e^2A^2/(2m)$] and in particular, the drive effect by electric field ${\bf E}$ are absent.  
Without the drive effect from ${\bf E}$, the Eilenberger equation can not reduce to the well-known Boltzmann equation for normal metals when the order parameter tends to zero from $T<T_c$ to $T>T_c$. More seriously, in the Eilenberger equation [Eq.~(10) in Ref.~\onlinecite{CO}], all electromagnetic effects vanish as the order parameter (i.e., anomalous Green function) becomes zero from superconducting to normal states. This apparently is unphysical.\cite{Kitafix} 

Secondly, the conclusion that the Higgs mode is not
sensitive to disorder in Ref.~\onlinecite{CO}, is incorrect. 
This can be easily
seen by the following simple analysis through the general
physics. In Nambu space,\cite{gi0,gi1} the Bogoliubov-de Gennes Hamiltonian in the
presence of the Higgs-mode excitation $\delta|\Delta|$ is written as $H_{\rm
  BdG}=\xi_{\hat p}\tau_3+\Delta_0({\bf r})\tau_1+\delta|\Delta({\bf
  r})|\tau_1$ in the real space, and the electron-impurity
interaction is given by $V({\bf r})\tau_3$. Then, due to the
non-commutation relation
\begin{equation}\label{RE1} 
  [\delta|\Delta({\bf 
      r})|\tau_1, V({\bf r})\tau_3]\ne0,
\end{equation}
the Higgs mode must be sensitive to the disorder.

In Ref.~\onlinecite{CO},  Silaev\cite{CO} claimed that the
impurity scattering of the isotropic part vanishes and then
concluded that the Higgs mode is insensitive to disorder.
Nevertheless, in both his diagrammatic formalism and Eilenberger
equation, the quasiclassical approximation with
an integration over the energy variable on Green function was applied.\cite{CO}
However,the scattering effect in the approach with this approximation is incomplete. Only the scattering influence on the anisotropic part of the Green function (i.e., anisotropic excitation by the paramagnetic effect) is included, whereas the influence on the isotropic part of the Green function which determines the Higgs mode is dropped out.

Specifically, the Eilenberger equation\cite{Eilen} originates from the basic Gorkov equation\cite{G1} by using $\tau_3$-Green function. In the Gorkov equation, the impurity self-energy of the nonequilibrium $\tau_3$-Green function reads\cite{G1}
\begin{equation}
  \Sigma_{\bf k}=n_i\int{d{\bf q}}|V_{\bf
    k-q}|^2{\delta}G(\omega,{\bf q}).\label{IS1}
\end{equation}
Due to Eq.~(\ref{IS1}), the  solved nonequilibrium Green function from the Gorkov equation is written as
\begin{equation}
{\delta}G=\delta{G}_{cl}+\delta{G}_{sc},
\end{equation}
which consists of clean-limit part $\delta{G}_{cl}$ and scattering correction $\delta{G}_{sc}$. It is noted that Eq.~(\ref{IS1}) is {\em finite} whether ${\delta}G(\omega,{\bf q})$ is isotropic or anisotropic in the momentum space. For the anomalous Green function $\delta{f}$ in ${\delta}G$, one finds a finite isotropic part in the off-diagonal component of $\Sigma_{\bf k}$:  
\begin{eqnarray}
  \Sigma_{k}|^{f}_{s}&\!=\!&n_i\int{d\Omega_{\bf q}}\int{d{\bf q}}|V_{\bf
    k-q}|^2\delta{f}(\omega,{\bf q})\nonumber\\
  &\!\approx\!&n_i\int{d\Omega_{\bf k}}\int{d{\bf q}}|V_{\bf
    k_F-q_F}|^2\delta{f}(\omega,{\bf q})\nonumber\\
&\!=\!&\Gamma_{0}\int{d\xi_q}\langle\delta{f}(\omega,{\bf q})\rangle,  
\end{eqnarray}
where {\small $\Gamma_{0}=n_iD\int{d\Omega_{\bf k}}|V_{\bf 
    k_F-q_F}|^2$} with $D$ representing the density of states and $\langle~\rangle$ denotes the angle average. This {\em finite} self-energy $\Sigma_{k}|^{f}_s$ 
determines a finite isotropic part of anomalous Green function $\delta{f_{sc}}(\omega,{\bf k})$ and hence leads to 
a nonzero scattering influence on the Higgs mode through the BCS gap equation.

In the derivation of the Eilenberger equation from
Gorkov equation,\cite{Eilen,Eilen1} the impurity collision integral comes from the
commutation between impurity self-energy [Eq.~(\ref{IS1})] and
Green function, i.e., $[\Sigma_{\bf k},G(\omega,{\bf k})]$, which is
finite even when $G(\omega,{\bf k})$ is isotropic in the momentum space. 
However, after the quasiclassical approximation with 
an integration over the energy variable
$g=\int{d\xi_k}{G(\omega,{\bf k})}$, the impurity collision
integral in the Eilenberger equation $\tau^{-1}_{\rm
  imp}[\langle{g}\rangle,g]$ is derived.\cite{CO,Eilen,Eilen1}
Unfortunately, with this approximation,
the impurity collision integral
of the isotropic part of the Green function
$g_s=\langle{g}\rangle$ vanishes
(i.e., $[\langle{g}\rangle,g_s]=0$). Therefore, it is clearly seen 
that the original finite scattering influence on the isotropic part in
the Gorkov equation is dropped out after the quasiclassical
approximation, making the Eilenberger equation inadequate to handle
this scattering effect.
The diagrammatic formalism, which is equivalent to the calculation
of the Gorkov equation,\cite{G1} should have found this problem.
Nevertheless, in Ref.~\onlinecite{CO}, Silaev followed the same
procedure as the derivation of the Eilenberger equation\cite{Eilen1} to handle the
diagrammatic formalism. 
He first took the commutation between the self-energy and Green
function and applied the quasiclassical approximation afterwards,
as seen in Sec.~IV~A in his paper. Then, he of course
recovered the results from Eilenberger equation, and 
missed the scattering influence on the isotropic part.

\begin{widetext}

\begin{appendix}

\section{Derivation of the Ginzburg-Landau equation from the Eilenberger equation}
\label{A_GC}

In this part, we present the derivation of the Ginzburg-Landau equation from the Eilenberger equation.\cite{Ba20} For the quasiclassical Green function $g(i\omega_n,{\bf R},{\hat {\bf k}})$ in Matsubara-frequency space, the Eilenberger equation in the clean limit reads
\begin{equation}
  [i{\omega_n}\tau_3-{\hat \Delta}({\bf R})\tau_3,g(i\omega_n,{\bf R},{\hat {\bf k}})]\!+\!i{\bf v}_F\cdot{\bm \nabla}_{\bf R}g(i\omega_n,{\bf R},{\hat {\bf k}})\!+\!{e{\bf A}\cdot{\bf v}_F}[\tau_3,g(i\omega_n,{\bf R},{\hat {\bf k}})]=0.
\end{equation}
From above equation, by only considering anomalous Green function, one has
\begin{equation}
  2{i\omega_n}{g}^{(n)}_{12}=-i{\bf v}_F\cdot(\partial_{\bf R}-2ie{\bf A}){g}^{(n-1)}_{12},
\end{equation}
where $g^{(n)}$ denotes the $n$-th order response in the Green function. With the equilibrium zero order $g^{(0)}_{12}=\frac{\Delta}{\sqrt{(i\omega_n)^2-|\Delta|^2}}$, by only keeping lowest two orders, the solution of the anomalous Green function reads 
\begin{equation}
g_{12}=g^{(0)}_{12}+g^{(1)}_{12}+g^{(2)}_{12}=g^{(0)}_{12}-i\frac{{\bf v}_F\cdot(\partial_{\bf R}-2ie{\bf A})}{2{i\omega_n}}{g}^{(0)}_{12}-\frac{[{\bf v}_F\cdot(\partial_{\bf R}-2ie{\bf A})]^2}{4(i\omega_n)^2}{g}^{(0)}_{12}.\label{Ei2GL_2}
\end{equation}
Substituting the above solution into the gap equation\cite{Usadel} 
\begin{equation}\label{Ei-EG}
\Delta({\bf R})={UN(0)}T\sum_{\omega_n>0}{\langle}g_{12}(i\omega_n,{\bf R},{\hat {\bf k}})\rangle,  
\end{equation}
near the critical temperature, one obtains
\begin{equation}\label{GGGG}
T\sum_{\omega_n>0}\Big[\frac{v_F^2(\partial_{\bf R}-2ie{\bf A})^2}{12(i\omega_n)^3}-\frac{|\Delta|^2}{2(i\omega_n)^3}\Big]\Delta+\Big[\frac{1}{UN(0)}-\sum_{\omega_n>0}\frac{1}{i\omega_n}\Big]\Delta=0,
\end{equation}
which after some arithmetics exactly recovers the Ginzburg-Landau equation\cite{Ba20,G1}
\begin{equation}
\Big[\frac{7\zeta(3)E_F}{12(\pi{T})^2}\frac{(\partial_{\bf R}-2ie{\bf A})^2}{2m}+\ln\Big(\frac{T_c}{T}\Big)-\frac{7\zeta(3)}{8({\pi}T)^2}{|\Delta|^2}\Big]\Delta=0.
\end{equation}
Then, it is clearly seen from Eq.~(\ref{GGGG}) that if the lower limit of the summation for the Matsubara frequency (i.e., $\omega_n>0$) is overlooked, the electromagnetic effect (i.e., first term on the left-hand side) vanishes. 

It is also noted that for the Keldysh Green function widely used in the literature,\cite{Keldysh1,Eilen1} the gap equation reads
\begin{equation}
  \Delta(R)=\frac{UN(0)}{2}\int^{\infty}_{-\infty}{dE}\tanh(\beta{E}/2){\langle}g_{12}(E,R,{\bf \hat k})\rangle,  
\end{equation}
which is equivalent to Eq.~(\ref{Ei-EG}) after the standard construction of the closed contour.

\end{appendix}  
\end{widetext}


\begin{thebibliography}{0}
\bibitem{Kitafix} The lost gauge invariance in the derivation of the Eilenberger equation\cite{Eilen} from the gauge-invariant Gorkov equation\cite{G1} by $\tau_3$-Green function arises from the gradient expansion within the quasiclassical approximation.\cite{Kita} To fix this problem, Kita\cite{Kita,Ba20} constructed a gauge-invariant Green function by introducing the Wilson line\cite{Wilson} and then derived a gauge-invariant Eilenberger equation, in which the complete electromagnetic effects are kept. Nevertheless, Silaev in Ref.~\onlinecite{CO} used the original Eilenberger equation rather than the fixed gauge-invariant one.
\end{thebibliography}
\end{document}